\documentclass[11pt]{article}
\usepackage{latexsym,array,theorem,mathrsfs,bm,float}
\usepackage{psfrag}
\usepackage{amsfonts,amsmath,amssymb,latexsym,array,afterpage,
theorem,mathrsfs,bm,float,epsfig,color,graphicx,tabularx,here,multirow}
\usepackage[dvipdfmx]{hyperref}

\usepackage{fancybox}

  \makeatletter
    
    \@addtoreset{equation}{section}
  \makeatother

\usepackage[top=30mm,bottom=35mm,left=25mm,right=25mm]{geometry}

\newcommand{\bea}{\begin{eqnarray}}
\newcommand{\ena}{\end{eqnarray}}
\newcommand{\beann}{\begin{eqnarray*}}
\newcommand{\enann}{\end{eqnarray*}}

\newcommand{\ma}[1]{\mbox{$\mathcal{#1}$}}

\newcommand{\calhR}[1]{\raisebox{2ex}{\tiny ({\em h})}\hspace{-0.8em}{\ma R}}
\newcommand{\pd}{\partial}
\newcommand{\BS}{\boldsymbol}
\newcommand{\MC}{\mathcal}
\newcommand{\MF}{\mathfrak}
\newcommand{\MB}{\mathbb}

\newcommand{\p}{\prime}

\newif\iffigure
\figurefalse
\figuretrue

\begin{document}

\title{\bf{
Energy spectrum of a quantum spacetime \\ with boundary}
}

\author{
\\
Shoichiro Miyashita\thanks{e-mail address : miyashita-at-gravity.phys.waseda.ac.jp} 
\\ \\ 
{\it Department of Physics, Waseda University, 
3-4-1 Okubo, Shinjuku, Tokyo 169-8555, Japan} 
\\ \\
}

\date{~}

\maketitle
\thisfancyput(13.cm,1cm){
{\fbox{WU-AP/1907/01}}}
\begin{abstract}
In this paper, I revisit the microcanonical partition function, or density of states (DOS), of general relativity. By using the minisuperspace path integral approximation, I directly calculate the $S^2 \times Disc$ topology sector of the DOS of a (quantum) spacetime with an $S^2\times \MB{R}$ Lorentzian boundary from the microcanonical path integral, in contrast with previous works in which DOSs are derived by inverse Laplace transformation from various canonical partition functions.  Although I found there always exists only one saddle point for any given boundary data, it does not always dominate the possible integration contours. There is another contribution to the path integral other than the saddle point. One of the obtained DOSs has behavior similar to that of the previous DOSs; that is, it exhibits exponential Bekenstein--Hawking entropy for the limited energy range $ (1-\sqrt{2/3}) <GE/R_{b}< (1+\sqrt{2/3})$, where energy $E$ is defined by the Brown--York quasi-local energy momentum tensor and $R_{b}$ is the radius of the boundary $S^2$. In that range, the DOS is dominated by the saddle point. However, for sufficiently high energy, where the saddle point no longer dominates, the entropy approaches a constant value, different from in the previous DOSs, which approach zero.
\end{abstract}

\clearpage

\tableofcontents

\clearpage

\section{Introduction}

In quantum gravity, gravity would thermalize. In order to define thermal equilibrium in quantum gravity, we need some observables that will reach or tend toward equilibrium. These must be (quasi-)local observables defined on the boundary of a quantum spacetime since there do not exist local observables in the bulk if topology change is one of the properties of quantum gravity. Related to this fact, it is now well-known that we can define an energy--momentum tensor of (quantum) gravity on a spacetime boundary quasi-locally \cite{BrownYork1}. Thermodynamical quantities, such as energy or pressure, are obtained from this tensor in gravitational thermodynamics.

The information of gravitational thermal states could be obtained statistical mechanically. This kind of approach was initiated by Gibbons and Hawking and first applied to asymptotically flat spacetimes \cite{GibbonsHawking}. They proposed that a certain kind of Euclidean path integral of gravity can be the canonical partition function of gravity. This function is given by summing over all Euclidean geometries with the boundary $S^2 \times S^1$, where the length of $S^1$ represents the inverse temperature at the corresponding Lorentzian boundary. By using this formulation, they derived the ``free energy of a black hole (BH)'' for an asymptotically flat spacetime at zero-loop order and reproduced the BH entropy--area relationship.
\footnote{
``BH'' also means ``Bekenstein--Hawking.'' \cite{Bekenstein,Hawking1}
} 
.
However, one (big) problem of their partition function is that it does not represent the true thermal states of an asymptotically flat spacetime since all BH states are unstable. Presumably, it indicates there are no thermal states for an asymptotically flat spacetime. Later, York derived a canonical partition function for a spacetime with a Lorentzian boundary of finite radius $S^2 \times \MB{R}$ \cite{York1}. 
\footnote{
Before the work by York, Hawking and Page \cite{HawkingPage} found that the canonical partition function of asymptotically AdS spacetime would be well-defined and a stable BH phase would appear at high temperature. The behavior of York's canonical partition function and theirs are very similar.
}
According to the partition function, the BH phase becomes thermal states and gravitational thermodynamical entropy in the BH phase equals the BH entropy. 

One assumption they made is that the integration (hyper)contour is such that only real Euclidean solutions dominantly contribute to the path integral.
However, as was shown by Gibbons, Hawking, and Perry \cite{GibbonsHawkingPerry}, the integration contour for Euclidean gravitational path integrals cannot be a trivial real contour. It must be genuinely complex, which will generally pick up some complex saddle-point geometries and not all real ones. Therefore, if we choose a contour such that $n$ complex saddle points contribute, the partition function at zero-loop order is written as   
\bea
Z\simeq \sum_{k=1}^{n}e^{-I^{E,os}_{k}}, \label{EQintro} 
\ena
where $I^{E,os}_{k}$ is the value of the action at the $k$-th complex Euclidean metric satisfying the Einstein equation. From this perspective, Halliwell and Louko reconsidered the canonical partition function of gravity \cite{HalliwellLouko1}. 
\footnote{
Their conclusion was that there are no infinite convergent contours that reproduce York's canonical partition function at the zero-loop level. As I will explain in Section 4, there have been some attempts to define a canonical partition function that shares the same properties as York's \cite{LoukoWhiting, MelmedWhiting}.
}

Taking the above facts into account, in this paper, I reconsider the microcanonical partition function, or density of states (DOS), of general relativity (GR) with an $S^2\times \MB{R}$ Lorentzian boundary, which has been investigated several times in the literature \cite{BradenWhitingYork, LoukoWhiting, MelmedWhiting}.
All of the previously obtained DOSs are derived by inverse Laplace transformation from various canonical partition functions. Since which canonical partition function is correct to obtain the DOS is not clear, I propose another possible DOS obtained by a different approach, namely, a microcanonical path integral \cite{BrownYork2}. As was shown in \cite{BCMMWY}, thermodynamical ensembles and the action functionals of gravity (and a boundary condition of the gravitational path integral) are closely related. The complete form of the microcanonical action functional of gravity and the corresponding path integral were proposed by Brown and York \cite{BrownYork2}. In the path integral, the boundary condition is chosen such that the energy (density) is held fixed, which is suitable for defining the DOS directly from the gravitational path integral. The advantage of this approach is that we can (almost) straightforwardly obtain a DOS without worrying about how to obtain the correct canonical partition function for inverse Laplace transformation. Of course, in this approach, there is ambiguity in how to choose an integration contour of the path integral. However, as we will see, there is only one contour of the lapse integral having the desired property.

The remainder of this paper is organized as follows. In section 2, I review the partition function of GR and minisuperspace approximation, which is used to approximate infinite degrees of freedom of gravity within finite degrees. In the section, I introduce the minisuperspace metric used in this paper and show how thermodynamical quantities are written in terms of these variables. In section 3, I apply the minisuperspace method to the microcanonical path integral and, with the saddle-point approximation, obtain a one-dimensional lapse integral. The obtained ``on-shell'' action is not a one-valued function on the complex lapse plane and is found to be defined on three sheets. I consider various contours and show its consequences. In the last section, I summarize the result and discuss its relationships to previous works and its implications. Finally, open problems are listed.

\section{Partition function of GR and the minisuperspace path integral method}
In this section, I review the basics of statistical treatment of gravitational thermodynamics and how to approximate the Euclidean path integral of GR in order for it to be manageable. 

\subsection{Partition function of GR}
\iffigure
\begin{figure}
\hspace{1.5cm}
\includegraphics[width=5.5cm]{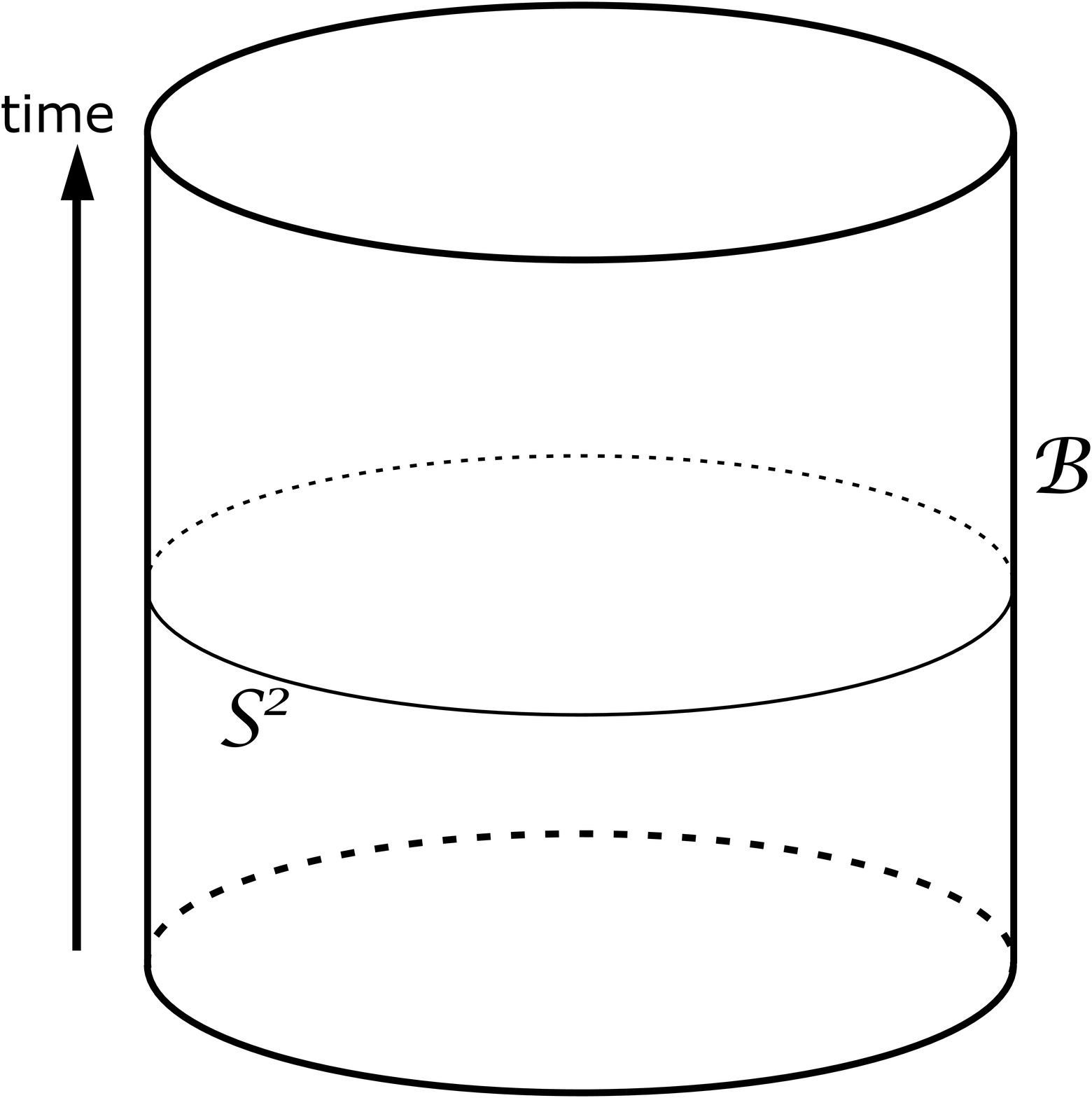}
\hspace{1.5cm} 
\includegraphics[width=6cm]{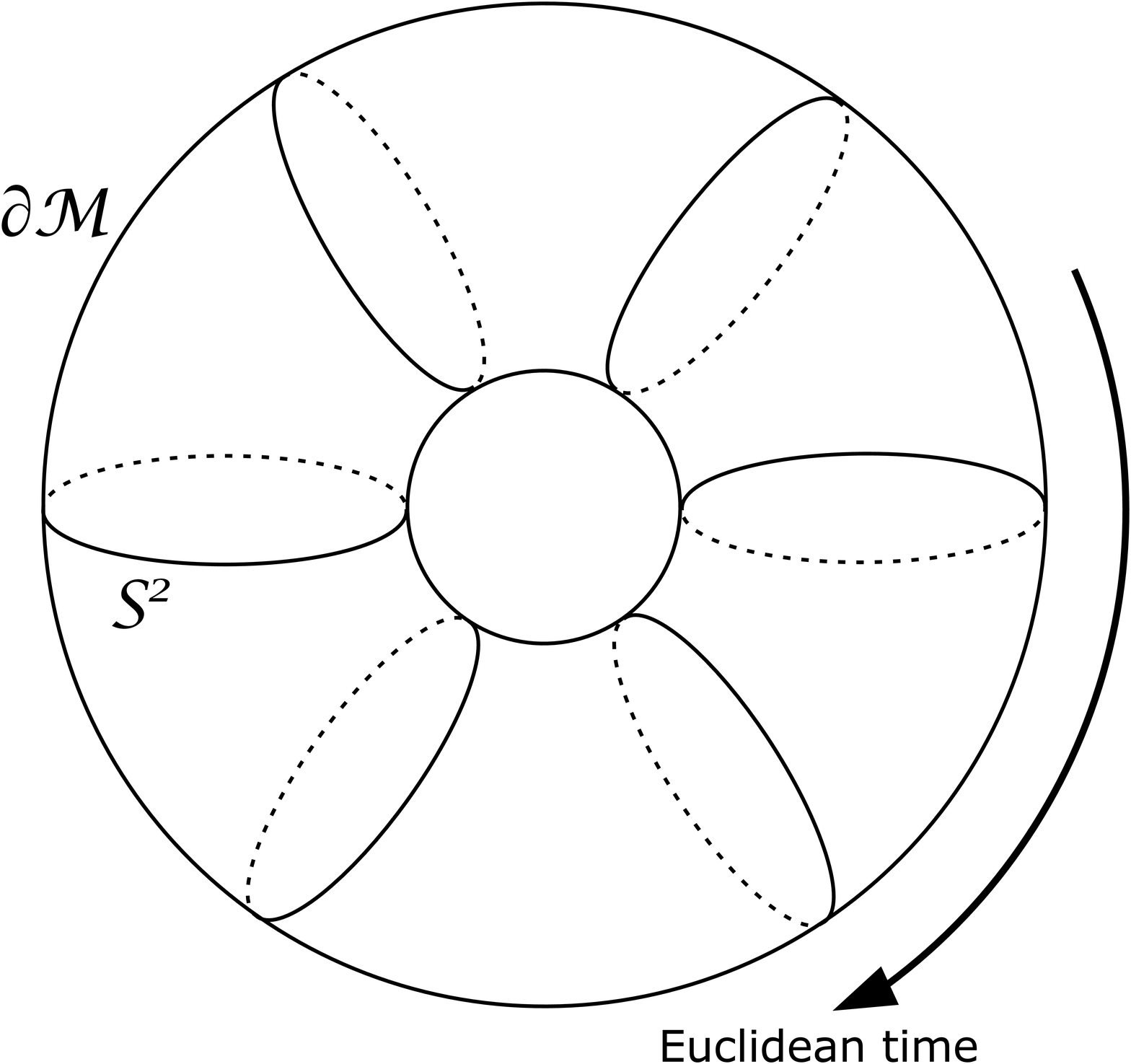}
\caption{LEFT: Spacetime with time-like boundary $\MC{B}=S^2 \times \MB{R}$. RIGHT: Euclidean geometry with $\pd \MC{M}=S^2 \times S^1$. }

\label{FIGst}
\end{figure}
\fi
In this paper, I consider thermal equilibrium states of a quantum spacetime with a Lorentzian boundary $\MC{B}= S^2 \times \MB{R}$ (Fig. \ref{FIGst}, left). When we say that {\it a  (quantum) spacetime is thermalized}, we mean that certain quantities defined on the boundary reach equilibrium values; that is, those quantities become almost isotropic and homogeneous on $S^2$ and time independent. As is usually done in statistical mechanics, the properties of such equilibrium states of gravity can be captured by the partition function of GR. This may be given by summing over all Euclidean histories with a boundary topology $\pd \MC{M}=S^{2} \times S^1 $ and satisfying suitable boundary conditions (Fig. \ref{FIGst}, right) \cite{GibbonsHawking, York1, BCMMWY}. Formally, 
\bea
Z_{\MF{E}}(\MF{Q},\MF{W})=\int_{\Gamma}\MC{D}\BS{g}_{\{ \MF{Q},\MF{W} \}} ~ e^{-I^{E}_{\MF{E}}[\BS{g}]}.  \label{EQpath}
\ena
I explain this equation below. \\
\\
$\blacksquare$ $\MF{E}$ represents ``Ensemble,'' where we can choose from the following: \\
~~~~~~~~ $\cdot$ microcanonical ensemble ($\MF{E}=mc$): Fixed energy and volume. \\
~~~~~~~~ $\cdot$ canonical ensemble ($\MF{E}=c$): Fixed temperature and volume.  \\
~~~~~~~~ $\cdot$ pressure microcanonical ensemble ($\MF{E}=pmc$): Fixed energy and pressure. \\
~~~~~~~~ $\cdot$ pressure canonical ensemble ($\MF{E}=pc$): Fixed temperature and pressure. \\
If we choose, for example, $\MF{E}=c$, then it represents the canonical partition function and we use the argument $\{\beta, V\}$, that is, $Z_c(\beta,V)=\int_{\Gamma} \MC{D}\BS{g}_{\{ \beta, V \}} e^{-I^E_{c}[\BS{g}]}$.\\
\\
$\blacksquare$ $I^{E}_{\MF{E}}[\BS{g}]$ is the Euclidean version of the action functional of GR;
\bea
I^{E}_{\MF{E}}[\BS{g}]=\frac{-1}{16\pi G} \int_{\MC{M}} d^4x \sqrt{g} \MC{R} + I^{E}_{\MF{E},\pd\MC{M}}[\BS{g}]
\ena 
where ensemble dependence is on only the boundary term $ I^{E}_{\MF{E},\pd\MC{M}}[\BS{g}]$. In general, suitable changes of boundary terms do not alter an equation of motion that gives the extrema of a path integral only when a type of boundary condition (hereinafter, ``BCtype'') is suitably chosen. In that sense, the choice of boundary term has one-to-one correspondence with choice of BCtype:
\beann
({\rm boundary ~ term } ) \longleftrightarrow ({\rm BCtype}).
\enann
The interesting thing is that, for the Euclidean gravitational path integral, certain BCtypes correspond to thermodynamical ensembles \cite{BCMMWY}.
This is because thermodynamical quantities are defined in reference to only the geometrical quantities on the spacetime boundary and because the conjugate pairs of thermodynamical quantities are indeed also the conjugate pairs in terms of BCtype
\footnote{
A pair of two quantities $X(y)$ and $Y(y)$ defined on the boundary is said to be a conjugate pair with respect to BCtype if there exist the following BCtypes: \\
~~~~~~ $\cdot$ fixing $X(y)$ while $Y(y)$ fluctuates, \\ 
~~~~~~ $\cdot$ fixing $Y(y)$ while $X(y)$ fluctuates, \\ 
and we can change one of the BCtypes to the other by adding or subtracting the term 
\beann
\int d^3 y X(y) Y(y).
\enann
}
. Therefore, the choice of boundary term corresponds to the choice of thermodynamical ensemble in quantum gravity:
\beann
({\rm boundary ~ term } ) \longleftrightarrow ({\rm BCtype}) \longleftrightarrow ({\rm thermodynamical ~ ensemble}).
\enann
In this way, thermodynamical ensembles and gravitation are closely related \cite{BCMMWY}.
For example, the boundary term for a canonical ensemble is the York--Gibbons--Hawking term \cite{York2, GibbonsHawking}, which is for a Dirichlet-type boundary condition. The reason why this is so for a canonical ensemble is that it fixes the thermal length $\beta$ of $S^1$ and the space volume $V$ of $S^2$ of the Euclidean boundary, and because the BCtype-conjugate quantities to $\beta$ and $V$ are the energy $E$ and pressure $P$, respectively (the definitions of $E$ and $P$ will be given later). The York--Gibbons--Hawking term is given by
\bea
I^{E}_{c,\pd \MC{M}}[\BS{g}]=\frac{-1}{8\pi G} \int_{\pd \MC{M}} d^3y \sqrt{\gamma} (\Theta-\Theta_{sub}(\BS{\gamma})), \label{EQYGH}
\ena
where $\BS{\gamma}$ is the induced metric on $\pd{M}$, $\Theta$ is the trace of the extrinsic curvature $\Theta_{\mu\nu}$ on $\pd \MC{M}$, and $\Theta_{sub}(\BS{\gamma})$ is the subtraction term in order for the on-shell action to be finite for asymptotically flat or AdS case. Note that we do not necessarily have to introduce it for the finite-radius $S^2 \times \MB{R}$ boundary case since the on-shell action is finite without subtraction in that case. As we will see shortly, it affects the definition of energy through the Brown--York tensor (\ref{defBYtensor}). Therefore, for convenience, I introduce the term in order for the energy of flat spacetime (enclosed by the $S^2\times \MB{R}$ Lorentzian boundary) to be zero. In particular, I take the background subtraction method of \cite{GibbonsHawking}, in which the subtraction term is given by the trace of the extrinsic curvature of flat spacetime $\Theta_{sub}=2/R_b$, where $R_{b}$ is the radius of the boundary $S^2$. For another example, the boundary term for microcanonical ensemble \cite{BrownYork2} is given by
\bea
I^{E}_{mc,\pd\MC{M}}[\BS{g}]=\frac{- 1}{8\pi G} \int_{\pd \MC{M}} d^3y \sqrt{\gamma} \tau_{\mu}\Theta^{\mu\nu}\pd_{\nu}\tau, \label{EQMCbdy}
\ena
where $\tau_{\mu}$ and $\tau$ represent a Euclidean time direction and a Euclidean time coordinate on $\pd \MC{M}$, respectively. Choosing this boundary term corresponds to choosing the BCtype fixing energy $E$ and allowing inverse temperature $\beta$ to fluctuate and fixing the space volume $V$.  Energy $E$ (and pressure $P$) is defined through the Brown--York quasi-local energy momentum tensor $\tau^{ij}$ \cite{BrownYork1}:
\bea
\tau^{ij}(y) \equiv \frac{2}{\sqrt{|\gamma(y)|}} \frac{\delta I_{c}[\BS{g}] }{\delta \gamma_{ij}(y)}= \frac{-1}{8\pi G} \left[\Theta^{ij}-\gamma^{ij} \Theta \right] -\frac{2}{\sqrt{|\gamma(y)|}}\frac{\delta I_{sub}[\BS{\gamma}]}{\delta \gamma_{ij}(y)}  ~ \label{defBYtensor},
\ena
where $I_{c}[\BS{g}]$ is the Lorentzian action corresponding to the canonical action and $I_{sub}[\BS{\gamma}]$ is the subtraction term, which is given by the negative of the second term of (\ref{EQYGH}). (The Euclidean version of the BY tensor is defined by just replacing $I_{c}[\BS{g}]$ with $I_{c}^{E}[\BS{g}]$ in the above equation.) 
In this way, energy and momentum (densities), and stress of the gravitational theory are defined on the boundary quasi-locally.
With this BY tensor, we can define total energy $E$ as
\bea
E \equiv \int_{S^2} d^2 z \sqrt{\sigma} u_{i}u_{j}\tau^{ij}, \label{EQEdef}  
\ena
where $\sigma_{ab} $ is the induced metric on the boundary $S^2$ and $u_{i}$ is normal to the boundary $S^2$ on $\MC{B}$.
In \cite{BrownYork2}, it was shown that choosing the boundary term (\ref{EQMCbdy}) and the BCtype that fixes energy $E$ (and volume $V$) leads to a well-posed variational problem
\footnote{
Precisely, fixing the energy density $u_{i}u_{j}\tau^{ij}$, the momentum density $-u_{i}\tau_{a}^{~i}$, and the stress tensor $\tau^{ab}$ leads to a well-posed variational problem. 
}
 and $E$ and $\beta$ form a conjugate pair with respect to BCtype.
Since the stress tensor (and energy density) become approximately isotropic, homogeneous on $S^2$, and time independent in gravitational thermal equilibrium, we can also define pressure $P$ as  
\bea
P(z) \equiv \frac{1}{2}\sigma^{ab}\tau_{ab} ~ .  \label{EQPdef} 
\ena
(Inverse temperature times) This pressure $P$ is also shown to be the BCtype conjugate to the volume $V$  \cite{BrownYork2} . 
Throughout this paper, I will consider the microcanonical action functional that consists of the Einstein--Hilbert term and this microcanonical boundary term. 
\\
\\
$\blacksquare$ $ \{ \MF{Q}, \MF{W} \}$, the subscript of $\MC{D}\BS{g}$, represents a boundary condition whose BCtype corresponds to the ensemble $\MF{E}$, in other words, the thermodynamical variables that are held fixed in the ensemble $\MF{E}$.
\\
\\
$\blacksquare$ $\MC{D}\BS{g}$ is the integration measure of the path integral. How it is defined is the one of the problems of the path integral of GR. In this paper, however, I will not go into detail about this and instead assume it does not affect the result of zero-loop approximation.  \\
\\
$\blacksquare$ $\Gamma$ represents the integration (hyper)contour of the path integral. Gibbons, Hawking, and Perry showed that the $\Gamma$ of the Euclidean path integral of GR must not be the real one in order to avoid the divergence problem \cite{GibbonsHawkingPerry}. Although we have to take some purely complex contour, we do not know which is the correct one. However, for the partition function, the contour must be chosen such that the path integral is real and positive valued. Additionally, following \cite{Hartle, HalliwellHartle}, I assume possible integration contours to be infinite (or closed). 

\subsection{Minisuperspace path integral method}
Evaluating the right-hand side of (\ref{EQpath}) is generally very difficult. Instead, an often-used method to capture its qualitative behavior is the minisuperspace path integral method, in which we truncate most of the degrees of freedom in the path integral. Throughout this paper, I concentrate on only the $S^2 \times Disc (D)$ topology and consider the following class of metrics:
\bea
\BS{g}=f(r)d\tau^2 + \frac{N^2}{f(r)}dr^2+R(r)^2 (d\theta^2 + \sin^2 \theta d\phi^2). \label{EQansatz} 
\ena
The coordinate variables $\theta \in (0,\pi)$, $\phi \in (0,2\pi)$ are the standard coordinate of $S^2$ and $\tau \in (0,2\pi)$, $r\in (0,1]$ is the polar coordinate of $Disc$, where $\tau$ represents angle and $r$ represents radius ($r=0$ corresponds the center and $r=1$ to the boundary)
\footnote{
Without loss of generality, the coordinate ranges of $\tau$ and $r$ can be set as done in the text since shifting and rescaling of the coordinate variables, together with suitable redefinition of $f$ and $N$, can lead to the form of the metric (\ref{EQansatz}) and the coordinate ranges $\tau \in (0,2\pi)$ and $r\in (0,1]$. I would like to thank an unknown referee who has provided valuable remarks on this point.}
.
Restricting the class of metrics summed over in the path integral to the form of the metric (\ref{EQansatz}), the partition function (\ref{EQpath}) is approximated to be
\bea
Z_{\MF{C}}(\MF{Q},\MF{W}) \simeq \int_{\Gamma} dN \MC{D}f \MC{D}R_{\{ \MF{Q},\MF{W} \}} e^{-I_{\MF{C}}^{E}[f,R;N]}.
\ena
However, this still be difficult to deal with. Following the usual method \cite{HalliwellLouko2}, I further simplify this integral by saddle-point approximation for $f$ and $R$:
\bea
Z_{\MF{C}}(\MF{Q},\MF{W}) \simeq \int_{\Gamma} dN e^{-I_{\MF{C}}^{E, os\{ \MF{Q},\MF{W} \}}(N)},
\ena
where $I_{\MF{C}}^{E, os\{ \MF{Q},\MF{W} \}}(N)$ is the ``on-shell'' action function of $N$ for the given boundary data $\{ \MF{Q},\MF{W} \}$.
Finally, the partition function reduces a one-dimensional complex integral along the contour $\Gamma$. 

In terms of the minisuperspace variables, energy (\ref{EQEdef}), volume, inverse temperature, and pressure (\ref{EQPdef}) can be written as
\bea
E= \left. \frac{R}{G}\left( 1-\sqrt{f}\frac{R^{\p}}{N} \right)\right|_{r=1} ~, \hspace{2.25cm} \\
V=4\pi R(1)^2 ~ , \hspace{4.45cm} \\
\beta= 2\pi \sqrt{f(1)} ~ , \hspace{4.35cm} \\
P=\left. \frac{1}{8\pi G} \left( \frac{\sqrt{f}}{N}\frac{R^{\p}}{R}+\frac{1}{2N}\frac{f^{\p}}{\sqrt{f}}-\frac{1}{R} \right) \right|_{r=1}.
\ena

\section{DOS of GR}
\subsection{Derivation of one-dimensional integral}
As I explained in the previous section, reducing the gravitational path integral to a one-dimensional integral is one simple way in order for it to be manageable. The first step is to derive the minisuperspace action functional for the microcanonical ensemble $I^{E}_{mc}[f,R;N]$. Since the full action for a microcanonical ensemble is
\bea
I_{mc}^{E}[\BS{g}]=\frac{-1}{16\pi G}\int_{\MC{M}} d^4 x \sqrt{g}\MC{R}+\frac{- 1}{8\pi G} \int_{\pd \MC{M}} d^3y \sqrt{\gamma} \tau_{\mu}\Theta^{\mu\nu}\pd_{\nu}\tau,
\ena
substituting the minisuperspace ansatz (\ref{EQansatz}) into the action leads to
\bea
I_{mc}^{E}[f,R;N] = \frac{-\pi}{G} \int^{1}_{0} dr \left[ \frac{ f (R^{\p})^2 }{N} +\frac{f^{\p}RR^{\p}}{N}+N \right] \hspace{4cm} \notag \\
 -\frac{\pi}{2 G}\left. \left( \frac{f^{\p}R^2}{N}+\frac{4f R R^{\p}}{N} \right) \right|_{r=0}+\left. \frac{2\pi}{G}\frac{RR^{\p}}{N}f \right|_{r=1} ~ , \label{EQactionfR} 
\ena
that is, the microcanonical partition function, or DOS, is now approximated to be 
\bea
Z(E,V) \simeq \int_{\Gamma} dN \MC{D}f \MC{D}R_{ \{E,V \} } e^{-I^{E}_{mc}[f,R;N]}. \label{EQPFfR} 
\ena
The next step is to derive the ``on-shell'' action function. In order to seek the saddle point for $f$ and $R$, take the variation with respect to $f$ and $R$:
\bea
\delta_{f,R} I^{E}_{mc}[f,R,N]= \frac{-\pi}{G} \int dr \left[ -\frac{RR^{\p\p} }{N}\delta f + \left( -\frac{2f R^{\p\p}}{N}-\frac{f^{\p\p}R}{N}-\frac{2f^{\p}R^{\p}}{N} \right)\delta R \right] \hspace{3cm} \notag \\
+ \left[ -2\pi\sqrt{f} ~ \delta\left\{ \frac{R}{G}\left(1-\sqrt{f}\frac{R^{\p}}{N}\right) \right\} + \frac{-\pi}{G}\left( \frac{2f R^{\p}}{N}+\frac{f^{\p}R}{N} -2\sqrt{f} \right)\delta R \right]_{r=1} \notag \\
-\frac{\pi}{2G}\left[ \frac{2RR^{\p}}{N}\delta f + R^2 \delta \left( \frac{f^{\p}}{N} \right) +\frac{4 f R}{N}\delta (R^{\p}) \right]_{r=0}. \hspace{4.5cm}
\ena
Since we are now considering the DOS $Z_{mc}(E,4\pi R_{b}^2)$, that is, a gravitational (minisuperspace) path integral with the boundary conditions
\bea
 \left. \frac{R}{G}\left(1-\sqrt{f}\frac{R^{\p}}{N}\right)\right|_{r=1} = E ~ , \label{EQbcon1}  \\
 R(1)=R_{b}, \label{EQbcon2} 
\ena
the variation at the boundary $r=1$ vanishes. Additionally, the smoothness of the metrics at the center of the $Disc$ requires the following conditions for all minisuperspace histories:
\footnote{
Here, I assume all the metrics summed over in the gravitational path integral are smooth. Another boundary condition that corresponds to summing over non-smooth metrics was proposed in \cite{LoukoWhiting}. 
}
\bea
f(0)=0 \label{EQccon1}  \\
\frac{1}{2}\frac{f'(0)}{N}= 1. \label{EQccon2} 
\ena
Therefore, the variation at the center $r=0$ also vanishes. These lead to the equations of motion for saddle-point geometries:
\bea
R^{\p\p}=0 \label{EQEOMf}   \\
f^{\p\p}R+2f^{\p}R^{\p} =0. \label{EQEOMR} 
\ena
The solution of these equations can be written as
\bea
R(r)=Ar+R_{H},  \\
f(r)=B-\frac{C}{R(r)},
\ena
where $A,R_{H}$,$B$, and $C$ are integration constants. From the boundary conditions (\ref{EQbcon2})--(\ref{EQccon2}), we obtain
\bea
f(r) = \frac{2N R_{H}}{R_{b}-R_{H}} \left(1-\frac{R_{H}}{R(r)}  \right) \label{EQfsol}, \\
R(r) = (R_{b}-R_{H}) r + R_{H} \label{EQRsol}.
\ena
Moreover, using the remaining boundary condition (\ref{EQbcon1}), we can derive the equation for $R_{H}$:
\bea
R_{H}(R_{b}-R_{H})^2-\frac{ N R_{b}}{2}\left(\frac{GE}{R_{b}}-1\right)^2=0. \label{EQforRH}
\ena
This indicates $R_{H}(N)$ is not a mere function but is a triple-valued function. Postponing a discussion about this multi-valuedness until the next subsection and using the saddle-point approximation with (\ref{EQfsol}) and (\ref{EQRsol}), we obtain
\bea
I_{mc}^{E,os\{E,4\pi R_{b}^2 \}}(N)
=\frac{-1}{G} \left[ - 2\pi R_{H}(N)(R_{b}-R_{H}(N)) + \pi N \right] -\frac{\pi R_{H}(N)^2}{G}. \label{EQactionN}
\ena
Finally, we obtain a one-dimensional complex integral expression of the DOS;
\bea
Z_{mc}(E,4\pi R_{b}^2) \simeq \int_{\Gamma}dN e^{-I_{mc}^{E,os\{E,4\pi R_{b}^2 \}}(N)} \label{EQoneintegral}
\ena
One thing I would like to note is that, if we do saddle-point approximation for $N$, we will obtain the expected expression
\footnote{
In the subscript of $N$, $\it{sp}$ denotes ``saddle point,'' not ``south pole.''
}
\bea
S(E,4\pi R_{b}^2)=\log Z_{mc}(E,4\pi R_{b}^2) \simeq \frac{\pi}{G}R_{H}(N_{sp})^2,
\ena
where I used the constraint equation
\bea
2\pi R_{H}(N)(R_{b}-R_{H}(N)) - \pi N=0,
\ena
which can be derived from the variation of (\ref{EQactionfR}) with respect to $N$.

\subsection{Riemann surface}

\iffigure
\begin{figure}[t]
  \begin{center}
    \begin{tabular}{c}
      \begin{minipage}{0.33\hsize}
        \begin{center}
\includegraphics[width=2cm]{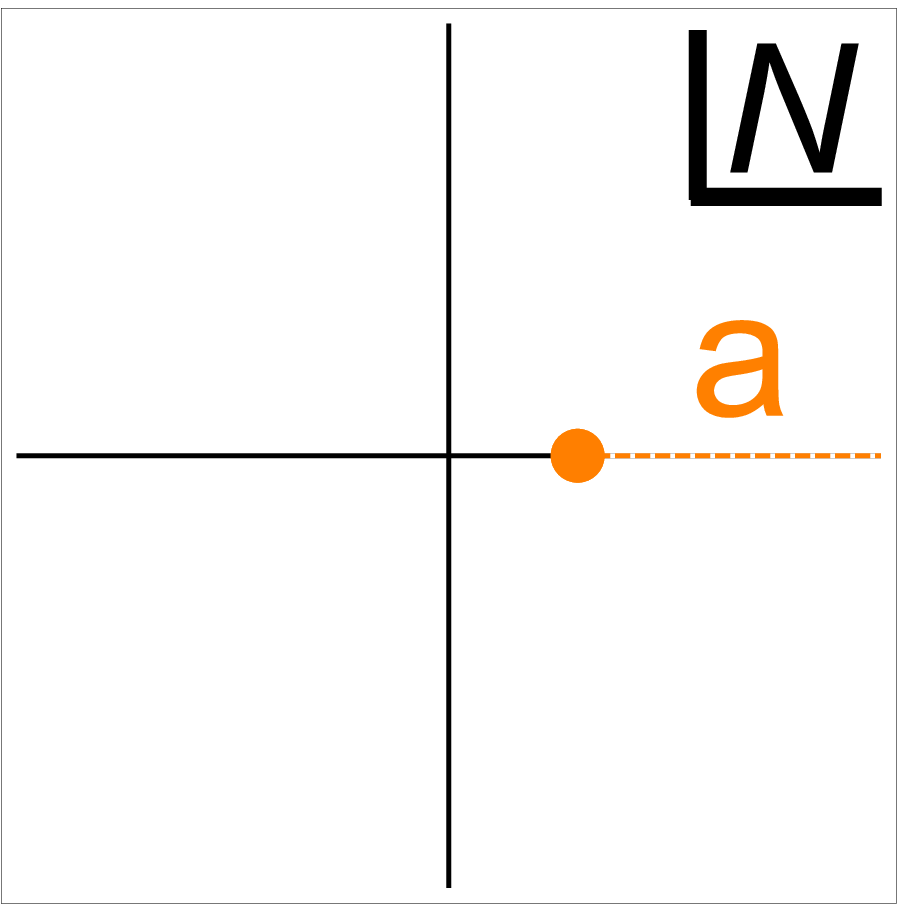} \\
\hspace{1cm} \\
\includegraphics[width=2cm]{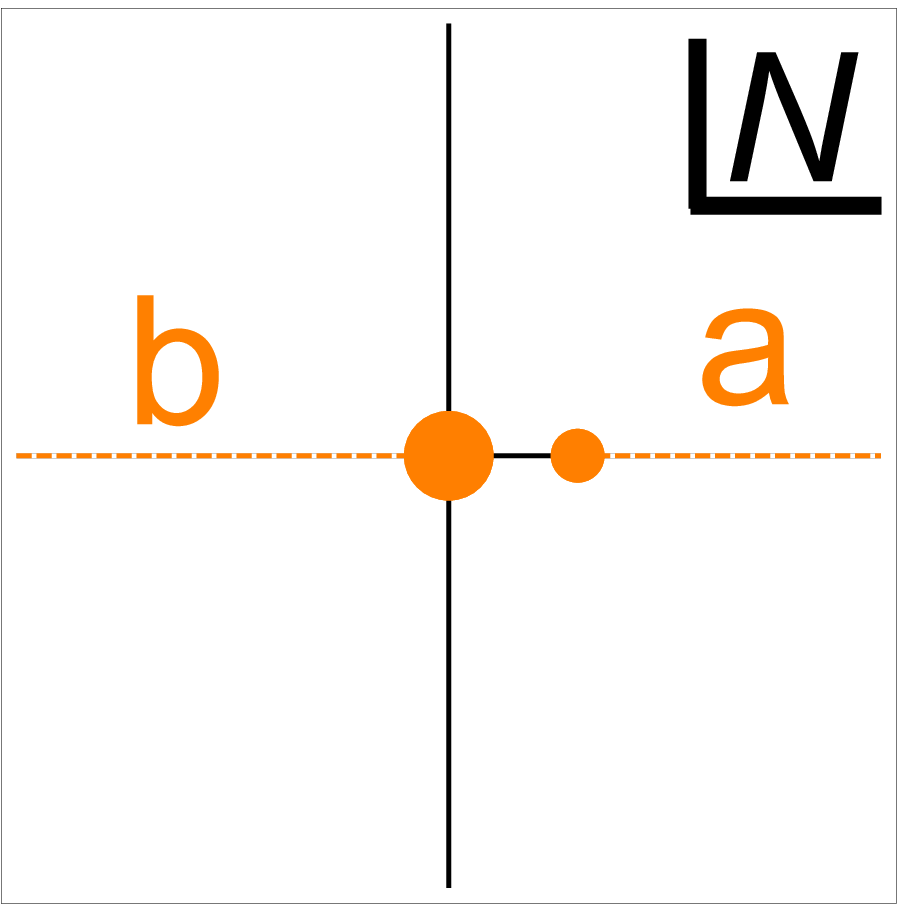} \\ 
\hspace{1cm} \\
\includegraphics[width=2cm]{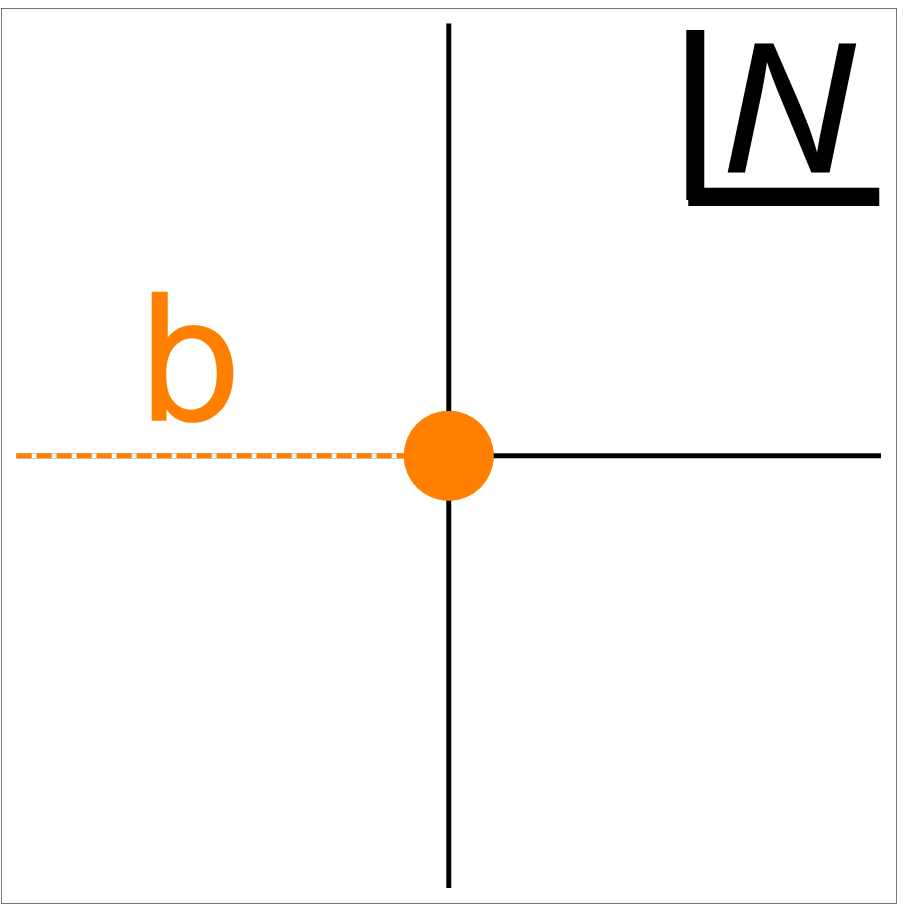} 
        \end{center}
      \end{minipage}
      \begin{minipage}{0.33\hsize}
        \begin{center}
          \includegraphics[clip, width=6cm]{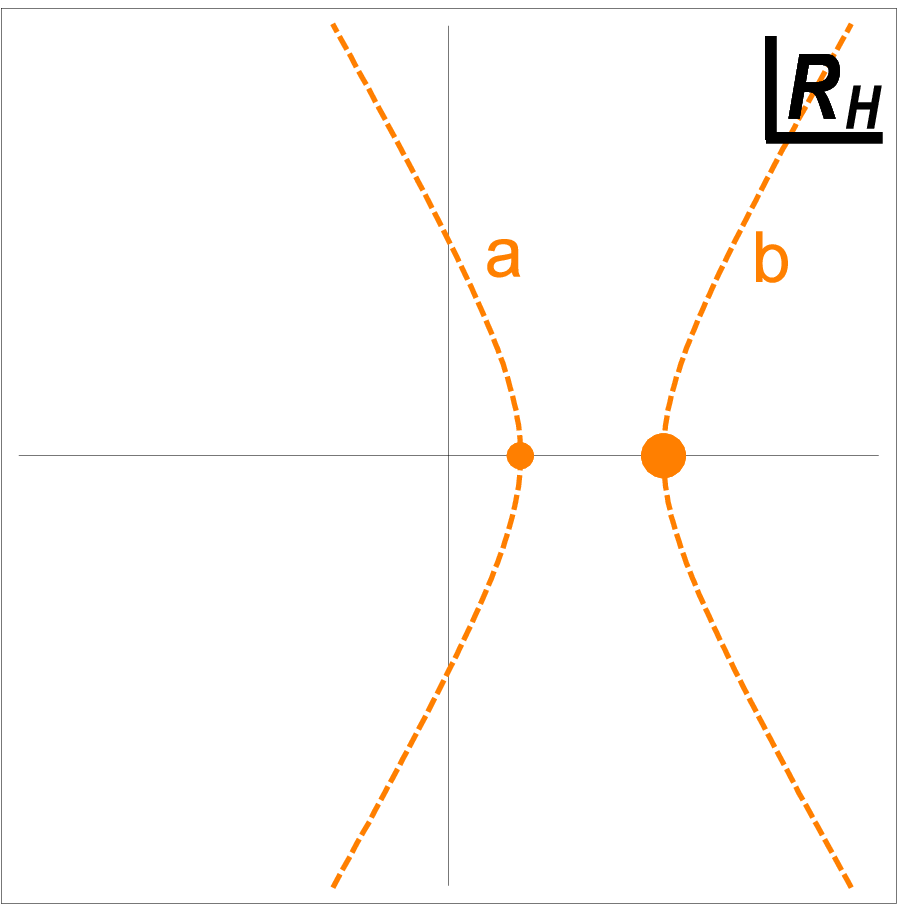}
        \end{center}
      \end{minipage}
      
    \end{tabular}
\caption{LEFT: Three complex $N$ planes consisting of the Riemann surface of the ``on-shell'' action. The orange circles represent the critical points of the map (\ref{EQN}). Orange dashed lines show one possible choice of branch cuts; branch cut $\BS{a}$ is $\left(\frac{8 R_{b}^2}{27\eta^2},\infty \right)$ on the real axis and branch cut $\BS{b}$ is $(-\infty,0)$ on the real axis.    RIGHT: An $R_{H}$ complex plane that is homeomorphic to the Riemann surface by the map (\ref{EQN}). The small circle represents $R_{H}=\frac{1}{3}R_{b}$ (corresponding to $N=\frac{8 R_{b}^2}{27\eta^2}$) and the large circle represents $R_{H}=R_{b}$ (corresponding to $N=0$).  }
\label{FIGA}
  \end{center}
\end{figure}
\fi

Since (\ref{EQactionN}) is not a single-valued function on a complex $N$ plane, we have to know what kind of Reimann surface the function (\ref{EQactionN}) is defined on. Since the inverse of $R_{H}(N)$ can be written as
\bea
N=\frac{2}{R_{b}\eta^2} R_{H}(R_{b}-R_{H})^2 \label{EQN}
\ena
by using (\ref{EQforRH}), the ``on-shell'' action function can be written as
\bea
I_{mc}^{E,os\{E,4\pi R_{b}^2 \}}(R_{H})
=\frac{\pi}{G} \left[ 2R_{b}\left(1-\frac{1}{\eta^2} \right)R_{H} + \left(\frac{4}{\eta^2}-3  \right)R_{H}^2 -\frac{2}{R_{b}\eta^2}R_{H}^3 \right]. \label{EQactionRH} 
\ena
Therefore, it can be a single-valued function on a complex $R_{H}$ plane. From the inverse map (\ref{EQN}), we can see there are two critical points, $R_{H}=R_{b}$ and $\frac{1}{3}R_{b}$, which correspond to $N=0$ and $\frac{8 R_{b}^2}{27\eta^2}$, respectively, on the $N$ planes. Then, three sheets are enough for (\ref{EQactionN}) and one example of two branch cuts as follows:
\beann
{\rm Branch ~ cut ~ }\BS{a}= \left\{N \left| Re(N)\in \left(\frac{8 R_{b}^2}{27\eta^2}, \infty  \right), Im(N)=0 \right. \right\} \\
{\rm Branch ~ cut ~ }\BS{b}= \left\{N \left| Re(N)\in \left(-\infty , 0  \right), Im(N)=0 \right. \right\}. \hspace{0.95cm}
\enann 
The upper sheet has the branch cut $\BS{a}$, the middle sheet has both branch cuts, and the lower sheet has the branch cut $\BS{b}$ (Fig. \ref{FIGA}, left).  The right panel of Fig. \ref{FIGA} shows the complex $R_{H}$ plane, on which I show the relevant region and how these regions correspond to the sheets.

\subsection{Integration contour}

\iffigure
\begin{figure}
\begin{center}
	\includegraphics[width=5.cm]{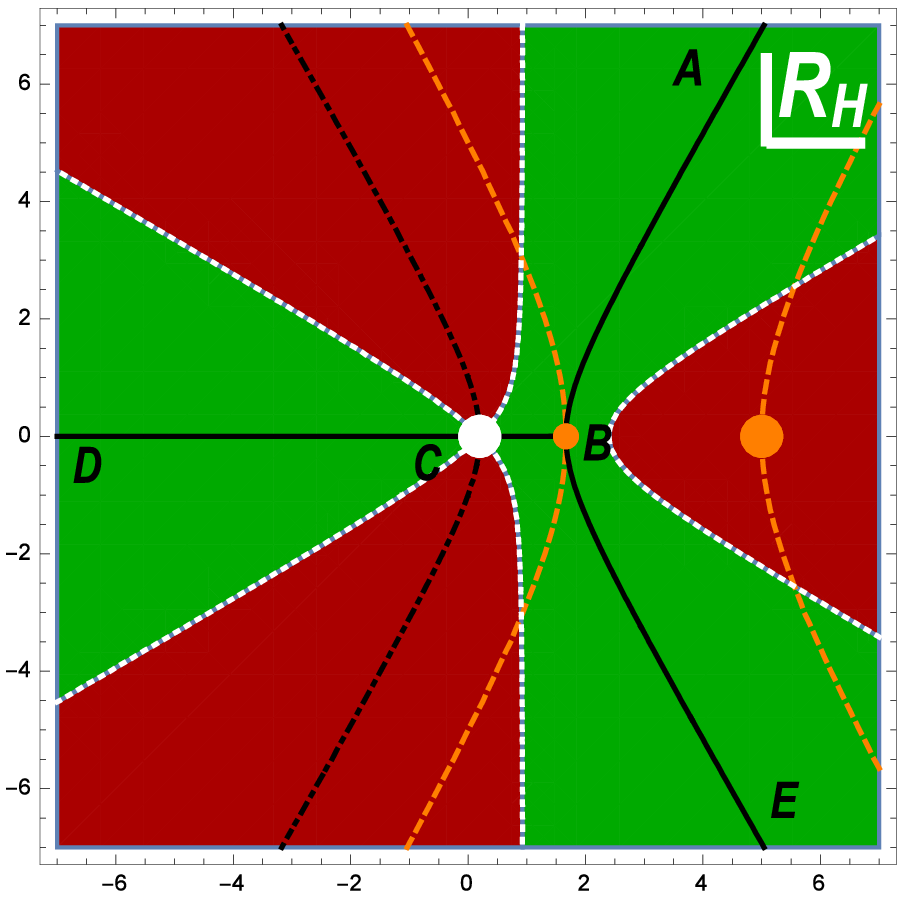} ~~
		\includegraphics[width=5.cm]{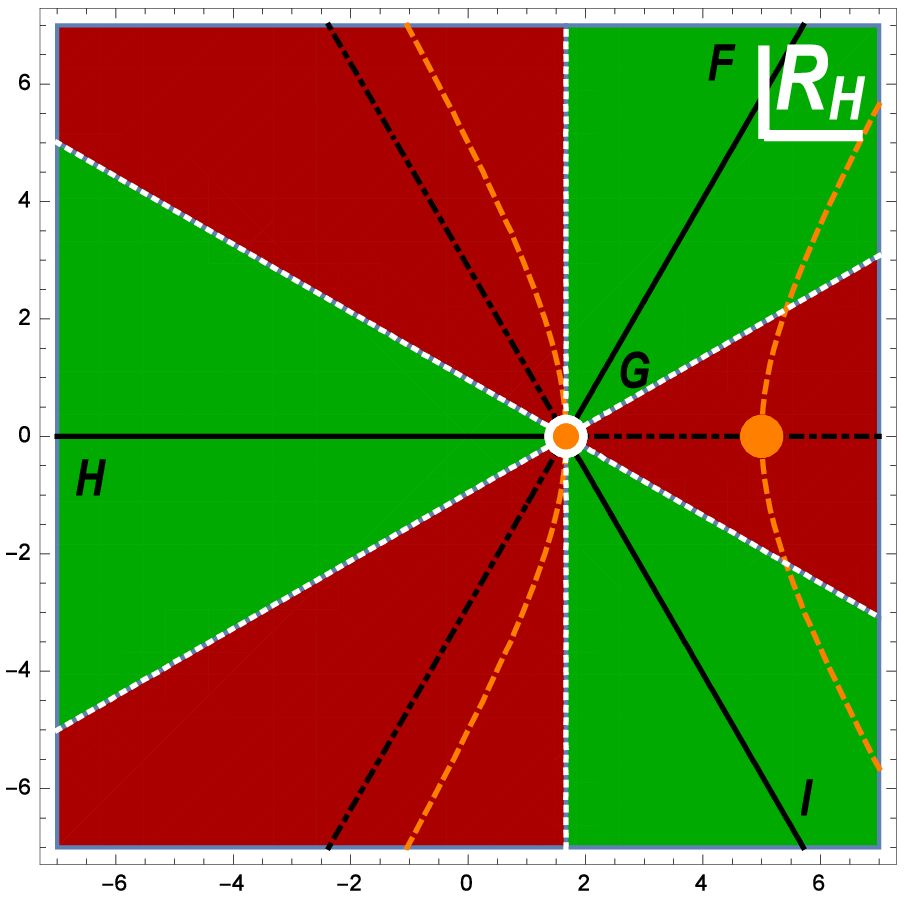} ~~
			\includegraphics[width=5.cm]{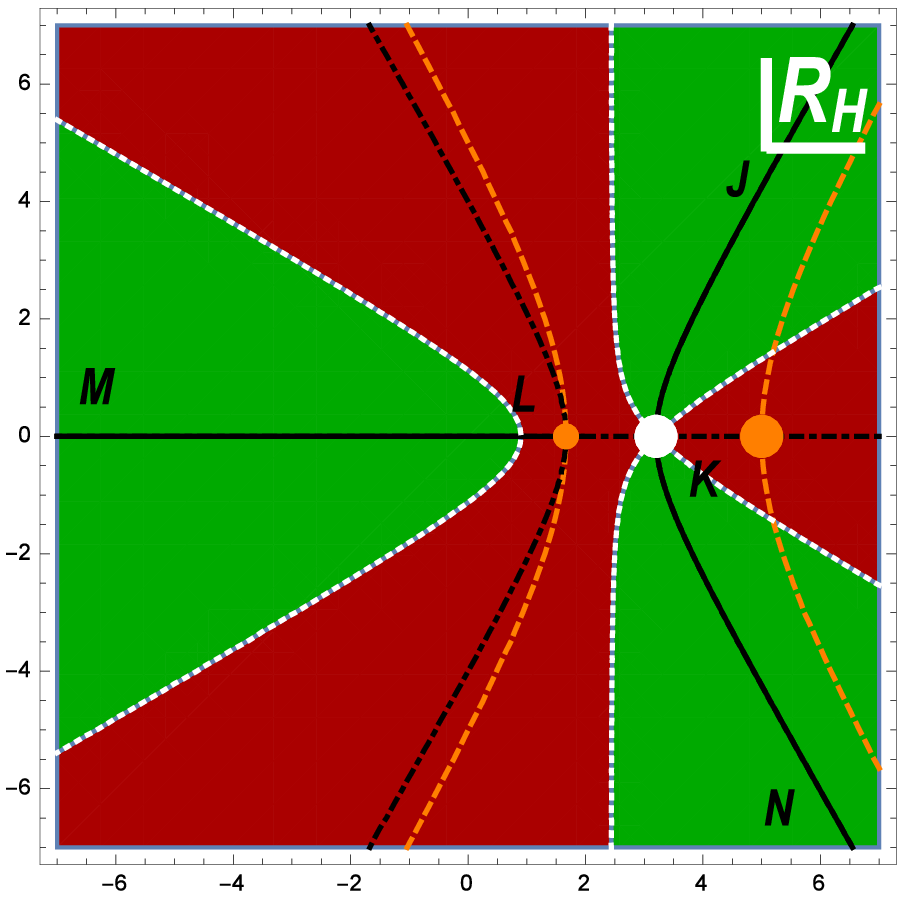}
	\caption{The location of the saddle point on the complex $R_{H}$ plane and its steepest descent (ascent) contours. The white circle represents the saddle point and solid (dot-dashed) black lines are its steepest decent (ascent) contours. The region where the real part of $-I^{E,os}_{mc}$ is higher (lower) than the saddle-point value is red (green) colored. LEFT: $R_{b}=5 \sqrt{G}, \sqrt{G}E=0.1 $ belonging to case (i). MIDDLE: $R_{b}=5 \sqrt{G} ,\sqrt{G}E=5\left( 1+\sqrt{\frac{2}{3}} \right)$ belonging to case (ii). RIGHT: $R_{b}=5\sqrt{G}, \sqrt{G}E=2$ belonging to case (iii). }
\label{FIGNcon}
\end{center}
\end{figure}
\fi

To estimate (\ref{EQoneintegral}), I will use the saddle-point approximation. From (\ref{EQactionRH}) and (\ref{EQN}), the location of the saddle point can be derived as follows: 
\bea
0= \left. \frac{\delta I_{mc}^{E, os\{E,4\pi R_{b}^2\}}(N)}{\delta N}\right|_{N=N_{sp}}= \frac{\pi}{G}\frac{\left( R_{H}(N_{sp})-(1-\eta^2)R_{b} \right)}{(R_{b}-R_{H}(N_{sp}))} \notag \\
\notag\\
\Longrightarrow ~~ R_{H}(N_{sp})=(1-\eta^2)R_{b}. \hspace{3cm}
\ena
For given $E$ and $R_{b}$, there is only one saddle point. Therefore, {\it if} we can choose the contour whose dominant contribution comes from only $N_{sp}$, then the DOS would be $\log Z_{mc} \simeq \frac{\pi}{G}R_{H}(N_{sp})^2$. However, as we will see shortly, there are no natural contours that give BH area-entropy relationships for any choice of $E$ and $R_{b}$. 

Depending on the energy $E$ and volume $4\pi R_b^2$, the behaviors of the ``on-shell'' action are qualitatively different and classified as follows:
\begin{center}
(i) $ 0<GE<R_{b} \left(1-\sqrt{2/3} \right)$ and $R_{b} \left(1+\sqrt{2/3} \right)<GE<\infty$; \\
(ii) $GE=R_{b} \left(1\pm \sqrt{2/3} \right)$; \\
(iii) $R_{b} \left(1-\sqrt{2/3} \right)<GE<R_{b}$ and $ R_{b}<GE<R_{b} \left(1+\sqrt{2/3} \right)$. \\
\end{center} 
Fig. \ref{FIGNcon} shows typical examples of (i)--(iii). Each figure is the complex $R_{H}$ plane, or equivalently, the three complex $N$ sheets, on which I show the saddle point (big white circle), the steepest descent and ascent contours for the saddle (solid black lines and dot-dashed black lines, respectively), and branch cuts (dashed orange lines). We can easily see that there exist essentially only two types of contour. In the case of (i), for example, these are ABE and A(E)BCD. We call them Type I and Type II, respectively.
\begin{center}
Type I : ABE for (i), ~~ FGI for (ii), ~~  JKN for (iii) ~~~~~~~~~~~~~~~~~~~ \\
Type II : A(E)BCD for (i), ~~ F(I)GH for (ii), ~~  J(N)KLM for (iii) \\
\end{center} 
In each case, there is a contribution other than the saddle point, that is, the points B and L of Fig. \ref{FIGNcon}. Since this is not the saddle point, we cannot apply saddle-point approximation. However, by changing the integration variable $N$ to $R_{H}$ by (\ref{EQN}), we find two ``saddle points'':
\bea
0= \left. \frac{\delta I_{mc}^{E, os\{E,4\pi R_{b}^2\}}(R_{H})}{\delta R_{H}}\right|_{R=R_{H,sp}}= -\frac{2 \pi}{G R_{b} \eta^2}(3R_{H,sp}-R_{b})(R_{H,sp}-(1-\eta^2)R_{b}) \notag  \\
\Longrightarrow ~~~~ R_{H,sp}=\frac{1}{3}R_{b},\ (1-\eta^2)R_{b}. \hspace{4cm}
\ena
In terms of $R_{H}$, (\ref{EQoneintegral}) is rewritten as
\bea
Z_{mc}(E,4\pi R_{b}^2) \simeq \frac{2}{R_{b}\eta^2} \int_{\Gamma} dR_{H} (R_{b}-R_{H})(R_{b}-3R_{H}) e^{-I_{mc}^{E,os\{ E,4\pi R_{b}^2 \}}(R_{H})}.\label{EQRHintegral} 
\ena
Similar to the method of \cite{BradenWhitingYork}, we could evaluate the contribution around the ``saddle points.''  At the zero-loop level, the partition function for each type is given by
\bea
Z^{I}_{mc}(E,4\pi R_{b}^2) \simeq
\begin{cases}
e^{\frac{\pi}{G}R_{b}^2 \frac{1}{3}\left(-1+\frac{8}{9\eta^2}  \right)} ~~~~~ {\rm for ~ (i)} \\
e^{\frac{\pi}{G}R_{b}^2(1-\eta^2)^2 } ~~~~~~~~~~ {\rm for ~ (ii) ~ and ~ (iii)} \label{EQsaddle1} 
\end{cases} \\
Z^{II}_{mc}(E,4\pi R_{b}^2) \simeq
\begin{cases}~
e^{\frac{\pi}{G}R_{b}^2(1-\eta^2)^2 } ~~~~~~~~ {\rm for ~ (i)} \\
e^{\frac{\pi}{G}R_{b}^2 \frac{1}{3}\left(-1+\frac{8}{9\eta^2}  \right)} ~~~~ {\rm for ~ (ii) ~ and ~ (iii)}. \label{EQsaddle2} 
\end{cases}
\ena
The behaviors of the corresponding entropies are shown in Fig. \ref{FIGen}.

\iffigure
\begin{figure}
\hspace{1.cm}
\includegraphics[width=7cm]{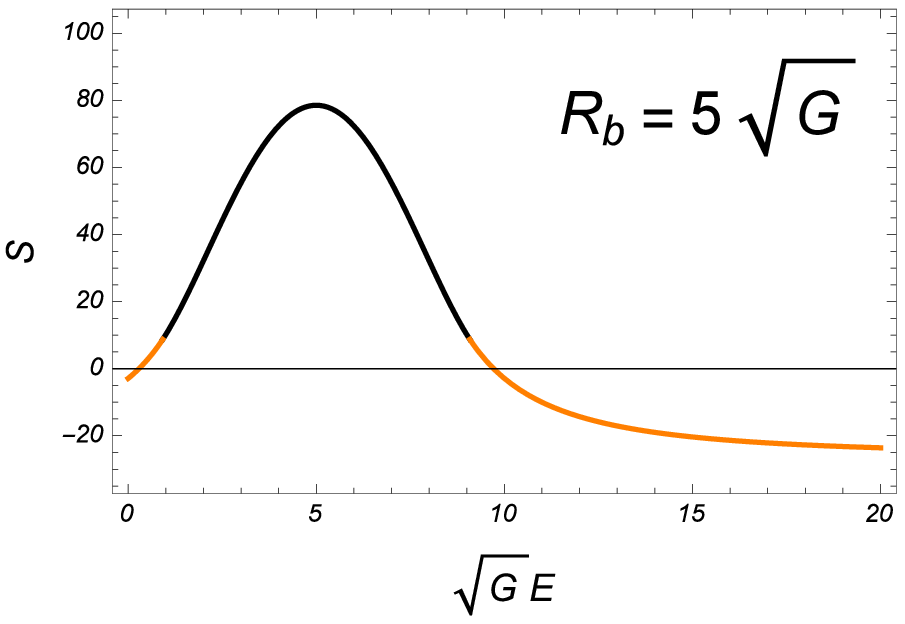}
\hspace{1.cm} 
\includegraphics[width=7cm]{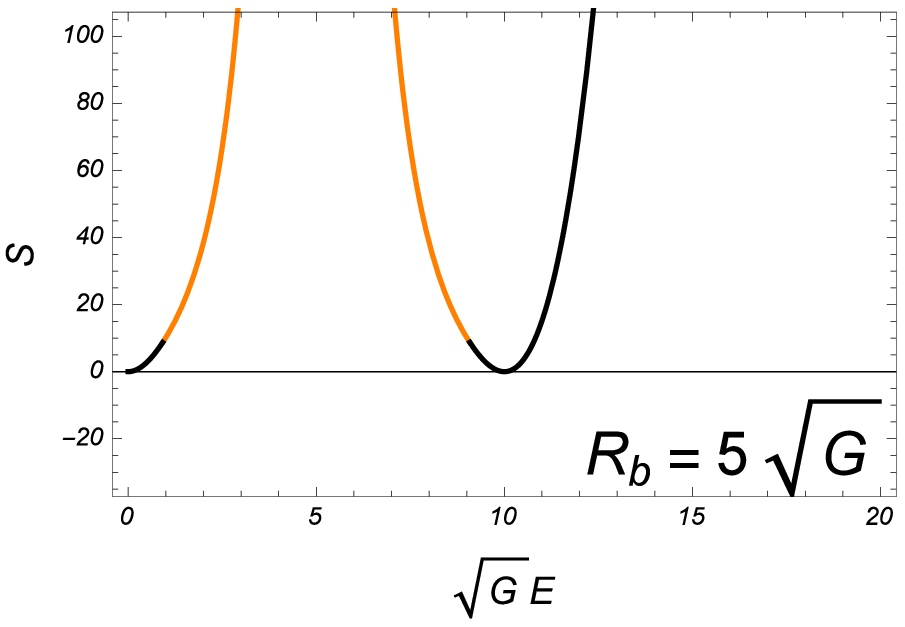}
\caption{The behavior of entropy corresponding to a type I DOS and type II DOS. Black curves indicate the contribution from the saddle point $R_{H}=(1-\eta^2)R_{b}$ and orange curves indicate the contribution from $R_{H}=R_{b}/3$. In each case, I set $R_{b}=5\sqrt{G}$. LEFT: type I DOS. RIGHT: type II DOS. }
\label{FIGen}
\end{figure}
\fi

\section{Discussion}
\subsection{Summary of the result}
In this paper, I evaluated the gravitational microcanonical partition function, or density of states (DOS), of an $S^2 \times \MB{R}$ Lorentzian boundary by using minisuperspace approximation and saddle-point approximation. I only considered the $S^2 \times D$ topology sector of the path integral. Following the conventional technique, I performed in advance the saddle-point approximation for minisuperspace functions $f$ and $R$ without specifying the hypercontour and obtained the one-dimensional lapse integral (\ref{EQoneintegral}). After that, I found there is only one saddle point of the ``on-shell'' action (\ref{EQactionN}) and showed that the Riemann surface of the ``on-shell'' action consists of three sheets. There are two ways to choose an infinite convergent contour. In each contour, there exists a contribution other than the saddle point. Taking this into account, I obtained two types of DOS (\ref{EQsaddle1}) and (\ref{EQsaddle2}) by the ``saddle point'' approximation
\beann
Z^{I}_{mc}(E,4\pi R_{b}^2) \simeq
\begin{cases}
e^{\frac{\pi}{G}R_{b}^2 \frac{1}{3}\left(-1+\frac{8}{9\eta^2}  \right)} ~~~~~ {\rm for ~} |\eta| \geq \sqrt{\frac{2}{3}} \\
e^{\frac{\pi}{G}R_{b}^2(1-\eta^2)^2 } ~~~~~~~~~~ {\rm for ~} |\eta| < \sqrt{\frac{2}{3}} 
\end{cases} \\
Z^{II}_{mc}(E,4\pi R_{b}^2) \simeq
\begin{cases}~
e^{\frac{\pi}{G}R_{b}^2(1-\eta^2)^2 } ~~~~~~~~ {\rm for ~} |\eta| \geq \sqrt{\frac{2}{3}} \\
e^{\frac{\pi}{G}R_{b}^2 \frac{1}{3}\left(-1+\frac{8}{9\eta^2}  \right)} ~~~~ {\rm for ~} |\eta| < \sqrt{\frac{2}{3}}, 
\end{cases}
\enann
where the ``shifted energy'' $\eta$ was defined by $\eta \equiv\frac{GE}{R_{b}}-1$. The behaviors of the corresponding entropies are shown in Fig. \ref{FIGen}.

\subsection{Relation to previous work}

As I mentioned in the introduction, there have been several attempts to derive the DOS of GR. In those works, inverse Laplace transformation (ILT) is used to obtain a DOS from a canonical partition function
\bea
Z_{mc}(E,V)=\frac{1}{2\pi i} \int^{i\infty +p}_{-i\infty+p} d\beta Z_{c}(\beta, V) e^{\beta E}  ~~~~ {\rm for ~ } p>0. \label{EQZmc} 
\ena
I will briefly explain these attempts and comment on the relationship (difference) among the results obtained in this work and in previous work. 
\subsubsection*{$\blacksquare$ ILT of York's canonical partition function \cite{BradenWhitingYork}} 
The first attempt to derive the DOS of a spacetime with finite radius $S^2 \times \MB{R}$ Lorentzian boundary is the work of Braden, Whiting, and York \cite{BradenWhitingYork}. In order to obtain the DOS, they considered the $S^2 \times D$ topology sector of York's canonical partition function $Z_{c}^{Y}(\beta, V)$ \cite{York1}
\footnote{
In \cite{York1}, York considered only the real saddle points of canonical Euclidean action and just assumed one of them gives the dominant contribution to the path integral without specifying its hypercontour.  
}
\footnote{
To be precise, since $Z_{c}^{Y}(\beta,4\pi R_{b}^2)$ must be a function of $\beta$ and $R_{b}$, it is given by
\beann
\log Z_{c}^{Y}(\beta, 4\pi R_{b}^2) \simeq \max [{\rm R.H.S. ~ of ~} (\ref{EQZc}) ],
\enann
and is defined for positive $\beta$ and $R_{b}$ that satisfy the condition $8\pi R_{b} \geq 3\sqrt{3}\beta$.
}
\bea
\log Z_{c}^{Y}(\beta, 4\pi R_{b}^2) \simeq -\frac{1}{G}\left[ 3\pi R_{H}(\beta,R_{b})^2 -4\pi R_{H}(\beta,R_{b}) R_{b}\left( 1-\sqrt{1-\frac{R_{H}(\beta,R_{b})}{R_{b}}} \right) \right], \label{EQZc} 
\ena
where the function $R_{H}(\beta,R_{b})$ is given by the equation
\bea
\frac{\beta}{4\pi}=R_{H}\sqrt{1-\frac{R_{H}}{R_{b}}}.
\ena
When we evaluate the integral (\ref{EQZmc}), $Z_{c}$ must be analytically continued (for $\beta$). As a result, they found the integral is defined on the Riemann surface, which consists of three complex $\beta$ sheets. They re-express ($\ref{EQZmc}$) with ($\ref{EQZc}$) as
\bea
Z_{mc}^{BWY}(E,4\pi R_{b}^2) \simeq \int_{\Gamma} d\xi (1+3\xi^2) \exp\left[ \frac{4\pi R_{b}^2}{G}\left( -\frac{1}{4}(3\xi^4+2\xi^2-1)-i \eta \xi (1+\xi^2) \right) \right],
\ena
where they defined $\xi$ by
\bea
\frac{\beta}{4\pi}=-i R_b \xi (1+\xi^2).
\ena
Their integration contour $\Gamma$ is uniquely determined by construction and convergence, and the dominant contribution comes from one of the three saddle points depending on $E$ and $R_{b}$: $\xi=-i\eta$ for $|\eta|<1/\sqrt{3}$ and $\xi=-i ~ {\rm sgn}(\eta)/\sqrt{3}$ for $|\eta|>1/\sqrt{3}$. Their DOS at the zero-loop level is 
\bea
Z^{BWY}_{mc}(E,4\pi R_{b}^2) \simeq
\begin{cases}
e^{\frac{\pi}{G}R_{b}^2 \frac{4}{3}\left(1-\frac{2}{\sqrt{3}}|\eta| \right)} ~~~~~ {\rm for ~ } |\eta| \geq \frac{1}{\sqrt{3}} \\
e^{\frac{\pi}{G}R_{b}^2(1-\eta^2)^2 } ~~~~~~~~~~ {\rm for ~ } |\eta| < \frac{1}{\sqrt{3}}. 
\end{cases} 
\ena
This is qualitatively similar to the type I DOS. Especially, for small $|\eta|$, both $Z^{BWY}_{mc}$ and $Z^{I}_{mc}$ give BH entropy. In terms of the microcanonical path integral, the contribution to their DOS, and also to the type I DOS, comes from the saddle-point geometry for small $|\eta|$ and not for large $|\eta|$. One of the differences between their formulation and the type I DOS is the transition point. The transition point $|\eta|=1/\sqrt{3}$ of $Z_{mc}^{BWY}$ is nothing but the critical geometry of York's canonical partition function where the stability changes.
\footnote{
To be precise, only $\eta=-1/\sqrt{3}$ corresponds to the critical geometry.
}
 On the other hand, the meaning of the transition point $|\eta|=\sqrt{2/3}$ of $Z^{I}_{mc}$ is not clear. Another difference is the behavior for large $|\eta|$. As I will comment in the next subsection, it is not clear whether the large $|\eta|$ behavior of $S^2 \times D$ topology sector is significant, If it is, the energy eigenstates may vanish at high energy in their formulation while they do not in type I case.

\subsubsection*{$\blacksquare$ ILT of Louko--Whiting canonical partition function \cite{LoukoWhiting}} 
After York's canonical partition function, Halliwell and Louko considered a canonical partition function in term of the minisuperspace (\ref{EQansatz}) and sought a suitable (infinite) integration contour on the complex $N$ plane that reproduce York's canonical partition function \cite{HalliwellLouko1}. However, their conclusion was negative. Following the result, Louko and Whiting considered a different canonical partition function by using different ``boundary'' conditions for the minisuperspace path integral. As was seen in Section 2, there are two boundaries for the minisuperspace path integral: $r=0$ and $r=1$. Of course, $r=0$ is not a true boundary. It is fictitious. The meanings of the boundary conditions (\ref{EQccon1}) and (\ref{EQccon2}) at $r=0$ are fixing the topology to be $S^2 \times D$ and ensuring the smoothness of the metrics at the center, respectively. They discarded the latter boundary condition, that is, they allowed non-smooth metrics to be summed over in a path integral. In order to discard the condition while maintaining the consistency of the variational principle, they added the ``boundary'' term $\left. \frac{\pi}{2G}R^2\left( \frac{f^{\p}}{N}-2 \right) \right|_{r=0}$ to the canonical minisuperspace action, which is straightforwardly obtained from the full action.
\footnote{
In the conventional path integral, where only smooth metrics are summed over, adding this term has no effect since this term equals zero for all Euclidean histories. However, when we extend the class of metrics summed over to include non-smooth metrics, it becomes very important as I will explain shortly.
}
It is given by  
\bea
I^{E,LW}_{c}[f,R;N]= \frac{-\pi}{ G} \int^{1}_{0} dr \left[ \frac{ f (R^{\p})^2 }{N} +\frac{f^{\p}RR^{\p}}{N}+N \right] \hspace{4cm} \notag \\
 -\left. \left( \frac{\pi R^2}{G} +\frac{2\pi f R R^{\p}}{G N} \right) \right|_{r=0}+\left. \frac{2\pi}{G} \sqrt{f}R \right|_{r=1}.
\ena
If we take the variation of their action with the boundary conditions $2\pi\sqrt{f(1)}=\beta$, $R(1)=R_{b}$, and $f(0)=0$, we would obtain the EOM for $f$ (\ref{EQEOMf}) for $R$ (\ref{EQEOMR}), the constraint equation, and the smoothness condition (\ref{EQccon2}) as the Euler--Lagrange equation for $R(0)$. One unsatisfactory point is that, as they remarked, the relationship to the full action is not clear. Since the ``boundary'' condition at $r=0$ is only fixing $f$, the natural expression for this minisuperspace path integral may be adding the integration of $R_{H}\equiv R(0)$. Ignoring possible non-trivial measures for the $R_{H}$ integral and performing the saddle-point approximation for $f$ and $R$, their path integral for the canonical partition function is reduced to the two-dimensional complex integral
\bea
Z_{c}^{LW}(\beta, 4\pi R_{b}^2) \simeq \int_{\Gamma} dR_{H} dN \exp \left[ \frac{1}{2G}\left[ \frac{1}{N}\beta^2 R_{b} (R_{b}-R_{H}) +N \right] +\frac{\pi}{G}R_{H}^2 -\frac{\beta R_{b}}{G} \right]. \label{EQcpLW} 
\ena
Their integration contour is the closed circle around the origin for $N$ and the finite interval $(0, R_{b})$ for $R_{H}$. This finite contour for $R_H$ is determined by the ``Wheeler--De Witt equation.'' 
\footnote{
Of course, since we are not considering wave functionals, there are no Wheeler--de Witt equations for partition functions. However, by construction, we can derive the differential equation that must be satisfied for gravitational partition functions, as we can derive a Wheeler--de Witt equation from the path integral expression of wave functionals \cite{HartleHawking}. They call this the ``Wheeler--de Witt equation'' for the partition function.
}
One remarkable point is that their canonical partition function shares some properties with York's while specifying a convergent integration contour explicitly. Then they obtained a DOS by ILT of the canonical partition function. At the zero-loop level, it is given by
\bea
Z^{LW}_{mc}(E,4\pi R_{b}^2) \simeq
\begin{cases}
0   \hspace{3.02cm} {\rm for ~ } |\eta| \geq 1 \\
e^{\frac{\pi}{G}R_{b}^2(1-\eta^2)^2 } ~~~~~~~~~~ {\rm for ~ } |\eta| < 1. 
\end{cases} 
\ena
Again, this DOS is similar to the type I DOS in the sense that it gives BH entropy for small $|\eta|$. The differences are, as before, at the transition point and the behavior with large $|\eta|$.

\subsubsection*{$\blacksquare$ ILT of Melmed--Whiting canonical partition function \cite{MelmedWhiting}} 
Melmed and Whiting again considered a canonical partition function with the same form as (\ref{EQcpLW}) but choosing a different integration contour \cite{MelmedWhiting}. After changing the integration variable in (\ref{EQcpLW}) by applying $\alpha\equiv N/(R_{b}-R_{H})$, they choose the integration contour to be the positive part of the real axis for $\alpha$ and the semi-infinite line parallel to the imaginary axis $  \left\{R_{H}=\frac{\alpha}{4\pi}+i k \left| k \in \left(-\sqrt{\frac{\alpha R_{b}}{2\pi}  }  \left(1-\frac{\beta}{\alpha} \right),\infty  \right) \right.  \right\}$. The resulting canonical partition function also shares some properties with York's. However, their DOS obtained from ILT of their canonical partition function is similar to, but differs from, Louko--Whiting's DOS:  
\bea
Z^{MW}_{mc}(E,4\pi R_{b}^2) \simeq
\begin{cases}
\frac{2}{R_{b}}\sqrt{\frac{G}{(\eta^2-1)}}   \hspace{1.4cm} {\rm for ~ } |\eta| \geq 1 \\
e^{\frac{\pi}{G}R_{b}^2(1-\eta^2)^2 } ~~~~~~~~~~ {\rm for ~ } |\eta| < 1 
\end{cases} 
\ena
at the leading order.
\footnote{
For $|\eta|\leq 1$, it is the zero-loop order. For $|\eta|\geq 1$, however, the zero-loop contribution is $1$ and the energy and volume dependence come from the evaluation including the neighborhood of the ``saddle point'' $\alpha=0$. 
}

\subsection{Further Remarks}
\begin{table}
\begin{center}
\begin{tabular}{| l || l | l | l |}
\hline 
 & large $|\eta|$ behavior & small $|\eta|$ behavior & transition point  \\
\hline \hline
$ Z^{BWY}_{mc} $ & $e^{\frac{\pi}{G}R_{b}^2 \frac{4}{3}\left(1-\frac{2}{\sqrt{3}}|\eta| \right)}$ & $e^{\frac{\pi}{G}R_{b}^2(1-\eta^2)^2 } $ & $|\eta|=\frac{1}{\sqrt{3}}$ \\
\hline 
$ Z^{LW}_{mc} $ & 0 & $e^{\frac{\pi}{G}R_{b}^2(1-\eta^2)^2 } $ & $|\eta|= 1$  \\
\hline 
$ Z^{MW}_{mc} $ & $ \frac{2}{R_{b}} \sqrt{\frac{G}{(\eta^2-1)}}$ & $e^{\frac{\pi}{G}R_{b}^2(1-\eta^2)^2 } $ & $|\eta|= 1$  \\
\hline 
$ Z^{I}_{mc} $ & $e^{\frac{\pi}{G}R_{b}^2 \frac{1}{3}\left(-1+\frac{8}{9\eta^2}  \right)}$ & $e^{\frac{\pi}{G}R_{b}^2(1-\eta^2)^2 } $ & $|\eta|= \sqrt{\frac{2}{3}} $  \\
\hline 
\end{tabular}
\caption{The list of the leading behavior of previously obtained DOSs and the type I DOS obtained in this work. All of these have transition points at finite $\eta$ and exhibit a BH entropy--area relationship for small $|\eta|$. They are evaluated at the zero-loop order except the large $|\eta|$ behavior of $Z^{MW}_{mc}$, which has vanishing entropy at the zero-loop order. }
\label{TA1}
\end{center}
\end{table}
As we saw in the previous subsection, there are many candidates for the ($ S^2 \times D$ topology sector of the) canonical partition function of GR, presumably due to the nonexistence of suitable infinite integration contours as shown by Halliwell and Louko \cite{HalliwellLouko1}. Since they were constructed in order to satisfy desired properties, such as the domination of the BH phase in the classical domain $E\lesssim R_{b}/G$, the DOSs obtained by ILT from them share the property that they reproduce BH entropy for small $|\eta|$. However, the small difference in the canonical partition functions results in differences of the high-energy (large $|\eta|$) behaviors of the DOSs. Therefore, in this work, instead of deriving the DOS by ILT from ambiguous canonical partition functions, I tried to derive the DOS directly from the microcanonical path integral. If the integration contour is supposed to be infinite, there are only two types of contour. One gives behavior similar to the previously obtained DOSs (type I) and the other has peculiar behavior (type II). I believe that the type I DOS describes the correct zero-loop behavior (of the $S^2 \times D$ sector) and the DOS will not vanish for arbitrarily high energy, in contrast with the previously obtained DOSs. Moreover, as I showed in section 3 (and as was also shown in \cite{MelmedWhiting}), the integration contours for all the DOSs listed in Table \ref{TA1} fail to capture the saddle point (i.e., the point satisfying the Einstein equation) for large $|\eta|$. If we think the gravitational path integral can always be approximated by the saddle-point(s)' contribution for any boundary data as (\ref{EQintro}), then the behavior for large $|\eta|$ must be replaced (dominated) by the other topology sectors, in which the path integration can be approximated by the saddle point(s), at least for large $|\eta|$. 

\subsection{Open Problems}
There are some open problems, which I list here. \\
$\cdot$ Finding the corresponding canonical partition function obtained from the Laplace transformation of type I (or II) DOS. \\
$\cdot$ The effect of the inclusion of matter fields or cosmological constants. \\
$\cdot$ Although there exists non-vanishing DOS for $E<0$, I ignored this in this work. Is it OK? \\
$\cdot$ As in the AdS case~\cite{Maldacena}\cite{Marolf}, for $0<E<R_{b}/G$, the purification of the microcanonical density matrix may correspond to an eternal BH geometry with boundary at $R_{b}$. What is the corresponding geometry for $R_{b}/G<E<2R_{b}/G$ (and for $2R_{b}/G < E$ if saddle-point geometries do not exist in the other topology sector)? Do there exist corresponding geometries for the purification for large $E$? \\
$\cdot$ The existence of saddle-point geometries in the other topology sectors which dominate the type I DOS for large $|\eta|$ should be explored. 

Some of these problems will be considered in \cite{Miyashita}

\section*{Acknowledgement}
S.M. is grateful to K. Maeda, S. Sato, and K. Okabayashi for useful discussions. 
This work was supported in part by a Grant-in-Aid (No. 18J11983) from the Scientific Research Fund of the Japan Society for the Promotion of Science.

\appendix

\appendix


\begin{thebibliography}{99}

\bibitem{BrownYork1} J. D. Brown and J. W. York, Jr. ,  \href{https://journals.aps.org/prd/abstract/10.1103/PhysRevD.47.1407}{Phys. Rev. D $\BS{47}$ (1993) 1407}
\bibitem{GibbonsHawking} G. W. Gibbons and S. W. Hawking , \href{http://journals.aps.org/prd/abstract/10.1103/PhysRevD.15.2752}{Phys. Rev. D $\BS{15}$ (1977) 2752}
\bibitem{Bekenstein} J. D. Bekenstein , \href{https://journals.aps.org/prd/abstract/10.1103/PhysRevD.7.2333}{Phys. Rev. D $\BS{7}$ (1973) 2333}
\bibitem{Hawking1} S. W. Hawking , \href{http://dx.doi.org/10.1007/BF01608497}{Commun. Math. Phys $\BS{43}$ (1975) 199}
\bibitem{York1} J. W. York, Jr , \href{https://journals.aps.org/prd/abstract/10.1103/PhysRevD.33.2092}{Phys. Rev. D $\BS{33}$ (1986) 2092}
\bibitem{HawkingPage} S. W. Hawking and D. N. Page , \href{http://dx.doi.org/10.1007/BF01208266}{Commun. Math. Phys. $\BS{87}$ (1983) 577}
\bibitem{GibbonsHawkingPerry} G. W. Gibbons, S. W. Hawking, and M. J. Perry , \href{http://dx.doi.org/10.1016/0550-3213(78)90161-X}{Nucl. Phys. B $\BS{138}$ (1978) 141}
\bibitem{HalliwellLouko1} J. J. Halliwell and J. Louko , \href{https://journals.aps.org/prd/abstract/10.1103/PhysRevD.42.3997}{Phys. Rev. D $\BS{42}$ (1990) 3997}
\bibitem{LoukoWhiting} J. Louko and B. F. Whiting , \href{http://dx.doi.org/10.1088/0264-9381/9/2/011}{Class. Quant. Grav. $\BS{9}$ (1992) 457}
\bibitem{MelmedWhiting} J. Memled and B. F. Whiting , \href{https://journals.aps.org/prd/abstract/10.1103/PhysRevD.49.907}{Phys. Rev. D $\BS{49}$ (1994) 907}
\bibitem{BradenWhitingYork} H. W. Braden, B. F. Whiting, and J. W. York, Jr. , \href{https://journals.aps.org/prd/abstract/10.1103/PhysRevD.36.3614}{Phys. Rev. D $\BS{36}$ (1987) 3614}
\bibitem{BrownYork2} J. D. Brown and J. W. York, Jr. , \href{http://journals.aps.org/prd/abstract/10.1103/PhysRevD.47.1420}{Phys. Rev. D $\BS{47}$ (1993) 1420}
\bibitem{BCMMWY} J. D. Brown, G. L. Comer, E. A. Martinez, J. Melmed, B. F. Whiting, and J. W. York, Jr. , \href{http://dx.doi.org/ 10.1088/0264-9381/7/8/020}{Class. Quant. Grav. $\BS{7}$ (1990) 1433}

\bibitem{York2} J. W. York, Jr , \href{https://journals.aps.org/prl/abstract/10.1103/PhysRevLett.28.1082}{Phys. Rev. Lett. $\BS{28}$ (1972) 1082}
\bibitem{Hartle} J. B. Hartle , \href{https://aip.scitation.org/doi/10.1063/1.528410}{J. Math. Phys. $\BS{30}$ (1989) 452}
\bibitem{HalliwellHartle} J. J. Halliwell and J. B. Hartle , \href{https://journals.aps.org/prd/abstract/10.1103/PhysRevD.41.1815}{Phys. Rev. D $\BS{41}$ (1990) 1815}
\bibitem{HalliwellLouko2} J. J. Halliwell and J. Louko , \href{https://journals.aps.org/prd/abstract/10.1103/PhysRevD.39.2206}{Phys. Rev. D $\BS{39}$ (1988) 2206}

\bibitem{HartleHawking} J. B. Hartle and S. W. Hawking , \href{https://journals.aps.org/prd/abstract/10.1103/PhysRevD.28.2960}{Phys. Rev. D $\BS{28}$ (1983) 2960}
\bibitem{Maldacena} J. Maldacena , \href{http://dx.doi.org/10.1088/1126-6708/2003/04/021}{JHEP $\BS{0304}$ (2003) 021}
\bibitem{Marolf} D. Marolf , \href{http://dx.doi.org/ 10.1007/JHEP09(2018)114}{JHEP $\BS{1809}$ (2018) 114}


\bibitem{Miyashita} S. Miyashita , ``Partition Functions for Gravitational Thermodynamics'', in preparation



\end{thebibliography}
\end{document}